\documentclass[aps,prb,twocolumn,superscriptaddress,showpacs]{revtex4}

\usepackage{graphicx}

\begin{document}


\title{Analysis of electron-positron momentum spectra of metallic
  alloys as supported by first-principles calculations}


\author{P. Folegati}
\email[Electronic address: ]{paola.folegati@polimi.it}
\affiliation{Dipartimento di Fisica e Centro L-NESS, Politecnico di Milano, Via
Anzani 42, 22100 Como, Italy}
\affiliation{Laboratory of Physics, Helsinki University of Technology, P.O. Box
1100, FI-02015 HUT, Finland}
\author{I. Makkonen}
\affiliation{Laboratory of Physics, Helsinki University of Technology, P.O. Box
1100, FI-02015 HUT, Finland}
\author{R. Ferragut}
\affiliation{Dipartimento di Fisica e Centro L-NESS, Politecnico di Milano, Via
Anzani 42, 22100 Como, Italy}
\author{M. J. Puska}
\affiliation{Laboratory of Physics, Helsinki University of Technology, P.O. Box
1100, FI-02015 HUT, Finland}


\date{\today}

\begin{abstract}
Electron-positron momentum distributions measured by the coincidence
Doppler broadening method can be used in the chemical analysis of
the annihilation environment, typically a vacancy-impurity complex in
a solid. In the present work we study possibilities for a quantitative
analysis, i.e., for distinguishing the average numbers of different
atomic species around the defect. First-principles electronic
structure calculations
determining self-consistently electron and positron densities 
and ion positions are performed for vacancy-solute complexes in 
Al-Cu, Al-Mg-Cu, and Al-Mg-Cu-Ag alloys. The ensuing simulated
coincidence Doppler broadening spectra are compared with measured ones
for defect identification. A linear fitting procedure, which uses
the spectra for positrons trapped at vacancies in pure 
constituent metals as components, has previously been employed to
find the relative percentages of different atomic species around 
the vacancy [A. Somoza \textit{et al}., Phys.\ Rev.\ B \textbf{65},
094107 (2002)]. We test the reliability of the procedure by the help of
first-principles results for vacancy-solute complexes and vacancies
in constituent metals.
\end{abstract}

\pacs{78.70.Bj, 81.40.Cd}

\maketitle

\section{\label{intro}Introduction}

In metallic alloys solute-vacancy association is one of the basic
processes controlling the precipitation phenomenon. Vacancies mediate
transport of solute atoms and contribute to the stability of precipitates
by reducing misfit stresses between them and the matrix 
material.~\cite{Ringer00} The precipitation process is technologically
important, since it improves mechanical properties of a number of light 
Al- and Mg-based alloys widely used in the vehicle industry.

Positron annihilation spectroscopy is a well-recognized technique for
investigating metallic alloys.~\cite{Dupasquier04,Dupasquier98} The
kinetics of precipitation after mechanical and thermal treatments can
be studied by positron lifetime spectroscopy while the chemical environment of
defects can be determined by Coincidence Doppler Broadening (CDB)
spectroscopy. The use of CDB spectroscopy was extended to study the
association of vacancies with solute atoms in metallic solid solutions
only a few years ago~\cite{Biasini01,Nagai00} but several CDB works
regarding alloys have been published since
then.~\cite{Somoza02,Dupasquier04,Dupasquier04b,Nagai01,Nagai02,Nagai04}

CDB spectroscopy suits especially well to study solute-vacancy complexes 
in Al alloys because, as open-volume defects, they trap effectively
positrons. Moreover, typical solute atoms, such as Cu, Zn and Ag, have
occupied $d$ atom shells giving strong signals in CDB momentum
spectra in contrast to Al. The analysis of CDB spectra is 
expected to give average concentrations of solute elements neighboring 
vacancies or vacancy-like defects at matrix-precipitate interfaces. Somoza et
al.~\cite{Somoza02}\ suggest to fit CDB spectra with
linear combinations of spectra measured for the matrix metal in the
annealed state and those of elemental samples corresponding to all the 
constituents of the alloy after severe  plastic deformation. The annealed 
matrix spectrum accounts for the annihilation of free positrons 
in bulk whereas the deformed sample spectra take into account the effects 
of positron confinement at vacancies. Recently, the scheme has been
revised by Ferragut~\cite{Ferragut06} to take into account the
non-saturation positron trapping by using the simultaneously 
measured positron lifetime spectra. Appendix shows examples 
of the fitting scheme in the case of Al-Cu alloy samples.

In the present work we study possibilities for quantitative analysis
of solute-vacancy complexes in metallic alloys. More specifically,
we choose the binary Al-Cu, ternary Al-Mg-Cu, and quarternary
Al-Mg-Cu-Ag alloys and model the solute-vacancy complexes in them
by first-principles electronic structure calculations. In the
simulations the ion positions are determined consistently with the 
electron density and also with the density of the trapped positron. 
The CDB spectra are also measured for the as-quenched
and one-minute aged Al--1.1 at.~\% Cu samples. The theoretical and
experimental spectra are compared in order to find the
most probable complexes present in the samples. The comparisons also 
emphasize the importance of fully relaxed ionic structures in 
calculating the CDB spectra. 

We also study the reliability of the fitting procedure
by Somoza \textit{et al.}\ by comparing the first-principles
results for different solute-vacancy complexes in binary, ternary,
and quarternary alloys with those obtained by the linear 
combination of spectra calculated for vacancies in elemental
deformed metals. A surprisingly good correspondence is found 
showing that the average numbers of solute atoms decorating the vacancy
can be determined. However, in some cases this may happen
with the price of modifying the fraction of trapped positrons.
Moreover, the distinction between adjacent atoms in the
Periodic Table, e.g.\ Mg and Al, may be difficult.

The present paper is structured as follows.
In Sec.~\ref{modeling} we describe the defect models used in this work
and in Sec.~\ref{method} our computational approach. Results and
discussion are presented in Sec.~\ref{results}. Finally, we present
our conclusions in Sec.~\ref{conclusions}.

\section{\label{modeling}Modeling}

In this work we  present the CDB data as ratio-difference (RD) spectra
\begin{equation}
\Delta(p)=\frac{\rho(p)-\rho_{\text{bulk}}(p)}{\rho_{\text{bulk}}(p)}, 
\end{equation}
where $\rho(p)$ and  $\rho_{\text{bulk}}(p)$ are
the spectra corresponding to the vacancy defect and perfect bulk systems,
respectively. This presentation is convenient, because partial 
annihilation in the bulk state (non-saturation trapping) in the
defect system does not affect the shape of the RD spectrum 
but only its amplitude.~\cite{Calloni05}
Moreover, in the ratio systematic theoretical errors in momentum 
density intensities cancel out.~\cite{Makkonen06}
When only one type of defect traps positrons, the trapping fraction
$F$ is
the multiplicative factor needed to scale the defect spectrum
(corresponding to saturation trapping) to coincide the measured
spectrum of the given defected system.~\cite{Calloni05}. This fraction can 
be compared
with the fraction obtained from positron lifetime measurements. 
However, usually in the case of alloys only mean lifetimes can be
extracted from the experiment and therefore $F$, in practice, acts as a
free parameter in the fitting procedure.

We start to build up the model for vacancy-solute defects in Al
alloys by considering deformed Al. This is because the positron data
for the alloy resembles that of deformed Al in the sense of
similar trapped positron lifetimes of 210\dots 220 ps 
(Ref.~\onlinecite{Marceau06}), which are
clearly shorter than the positron lifetime of 235 ps for as-quenched 
Al (Ref.~\onlinecite{Calloni05}).
In the previous work~\cite{Calloni05} we modeled a vacancy 
in deformed Al by moving the nearest-neighbor atoms of the vacancy
inward and keeping the more distant atoms in their ideal Al lattice
positions. The inward relaxation mimics the effects of strain 
due to dislocations with which vacancies are associated. In this
work we want to calculate from first-principles the positions of the 
Al and solute atoms and therefore the strain is taken into account by
reducing the Al lattice constant. We choose 
the strain to reproduce the experimental positron lifetime value 
of 225$\pm$1~ps for trapped positrons (Ref.~\onlinecite{Calloni05}). 
The strain needed is 4\% with respect to the experimental lattice 
constant of 4.05~\AA, resulting in a lattice constant of 3.89~\AA. 
Also the Doppler spectra for the vacancy in this strained Al 
is in a reasonable agreement with the experimental data for deformed Al
(Ref.~\onlinecite{Calloni05}). Namely, the predicted curve has the
correct shape and the scaling needed corresponds to the trapping
fraction of 48\% which is comparable to the value of 64\% 
deduced from the positron lifetime measurements.
Apart from using the empirically determined lattice constant we model
the geometries of the vacancy-solute complexes from first principles. This
means that the ionic structures of complexes are determined
by minimizing the Hellman--Feynman forces due to the electronic structure
and the localized positron state.~\cite{Makkonen06}

Here we want to remark that the determination of the amount of strain
is actually based not only on the agreement of the theoretical and
experimental positron lifetimes for the deformed Al; Also for
the unstrained case our lifetimes agree well with the experimental ones
(See Table~\ref{lifetimes}). Namely, 
for the bulk we get the theoretical lifetime of 165~ps
in good agreement with the experimental values of 153, 158, and
163~ps (Refs.~\onlinecite{Petters98,Staab96}
and~\onlinecite{Schaefer86b}) and our theoretical lifetime of 248~ps
for the vacancy is consistent with the experimental ones
of 235 \dots 248 ps
(Refs.~\onlinecite{Calloni05}~and~\onlinecite{Schaefer86b}). For the
vacancy-Cu complexes in Al the experimental
lifetimes are around 210~ps (Ref.~\onlinecite{Marceau06}) and our
modeling, as will be discussed below in more detail, gives typically
210\dots 220~ps.

In our tests for the fitting procedure by
Somoza \textit{et al.}\ we need also the CDB spectra corresponding to
vacancies in deformed Cu, Mg, and Ag. Similarly to the Al case their modeling
is performed by using slightly reduced lattice constants. They are 
given in Table~\ref{lifetimes} along with the resulting positron
lifetimes. The lifetimes are also compared with the measured ones.
For Cu and Ag the absolute values are typically too low, but the important
relative differences between the vacancy and bulk lifetimes agree well with 
experiment. In the case of Mg, theory predicts a larger lifetime difference 
than the measured one.

\begin{table}
\caption{\label{lifetimes} Theoretical $\tau_{\text{theor}}$ and experimental
 $\tau_{\text{exp}}$ positron lifetimes for elemental bulk metals and
 vacancies in them. Also the lattice parameters, $a$ for the
 face-centered cubic and the $c/a$ ratios for the hexagonal
 close-packed structures, are shown. The non-strained lattice
 parameters are experimental values at room temperature.~\cite{Kittel}}
\begin{ruledtabular}
  \begin{tabular}{lrrcrrr}
System & $a$ (\AA) & $c/a$ & $\tau_{\text{theor}}$ (ps) & $\tau_{\text{exp}}$ (ps)\\
\hline
Al bulk, no strain &  4.05 & & 165 &
153\footnote{Reference~\onlinecite{Petters98}}, 158\footnote{Reference~\onlinecite{Staab96}}, 163\footnote{Reference~\onlinecite{Schaefer86b}}\\
Al vacancy        &  & & 249 &
235\footnote{Reference~\onlinecite{Calloni05}},
236\footnotemark[1], 248\footnotemark[3]\\
Al bulk, strained  &  3.89 & & 149 & \\
Al vacancy, strained  & & & 226 & 225\footnote{Reference~\onlinecite{Folegati06}}, 240\footnote{Reference~\onlinecite{Staab00}}\\
Cu bulk, no strain & 3.61 & & 103 & 109\footnotemark[1], 112\footnotemark[2], 119\footnote{Reference~\onlinecite{Smedskjaer77}}\\
Cu vacancy & & & 175 & 168\footnotemark[1], 170\footnote{Reference~\onlinecite{Staab98}}, 180\footnote{Reference~\onlinecite{Dlubek87}}\\
Cu bulk, strained  & 3.52 & & 95 & \\
Cu vacancy, strained  & & & 160 & 170\footnotemark[8], 178\footnotemark[5]\\
Mg bulk, no strain & 3.21 & 1.62 & 230 &
219\footnote{Reference~\onlinecite{Ferragut06}}, 225\footnote{Reference~\onlinecite{Hautojarvi82}}\\
Mg bulk, strained  & 3.05 & 1.62 & 217 & \\
Mg vacancy, strained & & & 285 & 244\footnotemark[5]\\
Ag bulk, no strain & 4.09 &   & 123 &
131\footnote{Reference~\onlinecite{Schaefer86}}, 131\footnotemark[1]\\
Ag vacancy & & & 196 & 207\footnotemark[1]\\
Ag bulk, strained  & 3.97 & & 111 & \\
Ag vacancy, strained  & & & 189 & 196\footnotemark[5]\\
\end{tabular}
\end{ruledtabular}
\end{table}

\begin{figure}
\includegraphics[width=0.9\columnwidth]{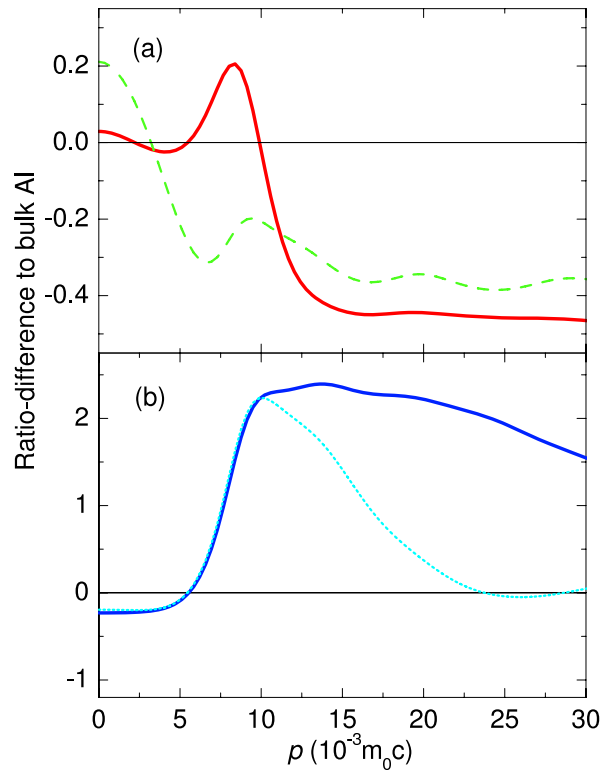}%
\caption{\label{RDelemental}(Color online) Calculated CDB RD spectra for 
  vacancies in elemental strained metals. (a) The solid (red) line
  shows the result for the vacancy in strained Al and the dashed
  (green) line for the vacancy in strained Mg. (b) The
  solid (dark blue) line is the result for a vacancy in strained Cu and the
  dotted (light blue) line for the vacancy in strained
  Ag.}
\end{figure}

Figure~\ref{RDelemental} shows the RD spectra relative to bulk Al calculated 
using the first-principles methods for vacancies in strained Al, Mg, Cu, and
Ag metals. At high momenta the Al and Mg RD spectra are rather similar,
reflecting the reduction of annihilation with similar electron cores. 
The Al spectrum has a peak around $9\times 10^{-3}\ m_{0}c$
due to positron confinement.~\cite{Calloni05} In the Mg RD
spectrum there is a maximum at low momenta reflecting the rather low
electron density (and long positron lifetime) at the vacancy. In the
Cu RD spectrum the high level at high momenta is due to annihilation
with Cu $3d$ electrons whereas in the case of Ag the peak at $10\times
10^{-3}\ m_{0}c$ originates from Ag $4d$ electrons. The difference
between the Cu and Ag RD spectra reflects the difference in the
localization of the $3d$ and $4d$ electrons. The latter are more
delocalized because $4d$ wave functions have to be orthogonal 
against $3d$ wave functions. The spectra in Fig.~\ref{RDelemental} are in
good agreement with the experimental data presented in
Ref.~\onlinecite{Ferragut06b} and with those used in the experimental
fitting procedure.~\cite{Ferragut06}

\section{\label{method}Computational method}

We model defects in the face-centered (fcc) Al, Cu, and Ag lattices 
using a cubic supercell of 32 atomic sites. For the hexagonal Mg
we use an orthorhombic supercell with 48 atomic sites.
Valence electron densities are computed self-consistently in the
framework of the local-density approximation (LDA) of the
density-functional theory; the computations are carried out by
employing the plane-wave code
\textsc{vasp}~\cite{Kresse96a,Kresse96b,PAWKresse} with the projector
augmented-wave (PAW) method~\cite{PAW} to account for the valence 
electron-ion core interaction. The first Brillouin zone of the 
superlattice is sampled using the $8\times 8\times 8$
Monkhorst-Pack $\mathbf{k}$-point mesh.~\cite{Monkhorst76} A plane-wave
cutoff of 342~eV is used when calculating the pseudo-valence wavefunctions, 
in which the rapid oscillations near to ion cores are smeared off.
The ionic structures of the vacancy-type defects are relaxed taking
into account also the forces on ions due to the localized
positron.~\cite{Makkonen06} Thereby electron and positron densities
and ion positions depend on each other and they are solved in a
self-consistent fashion. The PAW method enables also the
post-construction of all-electron valence wavefunctions which we use
to calculate accurately
electron momentum densities.~\cite{Makkonen05,Makkonen06} 
The cutoff equivalent to the momentum of the electron-positron pair of 
$70\times 10^{-3}\ m_{0}c$ is used when expanding all-electron valence 
wavefunctions in plane waves. 

Using the PAW total charge density including the free atom core 
electrons the positron potential is constructed within 
the LDA (Ref.~\onlinecite{Boronski86}) for electron-positron
correlation effects. Then the lowest-energy positron state
is calculated on a three-dimensional real-space point
grid.~\cite{Makkonen06} The so-called ``conventional scheme'', in which
the localized positron density does not affect directly the 
average electron density, is used to describe
trapped positrons. The annihilation rates of self-consistent
all-electron valence states and those of the atomic core electron states are
calculated within the LDA (Ref.~\onlinecite{Boronski86}) for the
electron enhancement at the positron. These partial annihilation rates are
used as weighting factors when calculating momentum densities of annihilating
electron-positron pairs within the so-called state-dependent
enhancement scheme.~\cite{Alatalo96,Barbiellini97} 
For Al matrix systems the momentum distributions corresponding to 
valence electrons are obtained by the three-dimensional Fourier
transformation on a cubic grid with the spacing of $0.67\times 10^{-3}\
m_{0}c$ and those for the core electrons on a dense radial grid using 
parameterized forms of the positron wavefunction.~\cite{Alatalo96} 
The momentum densities for bulk Cu, Ag, and Mg and vacancies in them
are calculated with accuracies similar to that for Al matrix. The ensuing
three-dimensional momentum distributions are integrated over planes
perpendicular to the [100] direction and convoluted with a Gaussian
function with the full width at half maximum (FWHM) of 
$3.6\times 10^{-3}\ m_{0}c$ in order to simulate experimental
CDB spectra. In the case of asymmetric defects we average over all
possible orientations of the defect. For a complete description of the
computational method see
Ref.~\onlinecite{Makkonen06}. Calloni \textit{et al}.~\cite{Calloni05}
have previously used the same methodology (without the first-principles
determination of the ionic relaxation) to study clean monovacancies
in strained Al.

\section{\label{results}Results and discussion}

\subsection{\label{structures}Ionic structures of vacancy-Cu complexes
in the Al-Cu alloy and positron lifetime results}


We have simulated vacancy-Cu complexes in the Al-Cu alloys by varying 
the number of Cu atoms, nearest neighbors to the vacancy, between 
1 and 12.
The resulting relaxed geometries and corresponding lifetimes are described
in Table~\ref{structuretable}. Up to 4 Cu atoms it is
reasonable to assume that the Cu atoms are on the same [100] plane
around the Al vacancy maximizing the Cu-Cu distances. This is because
according to Atomic Probe Field Ion Microscopy (APFIM) experiments Cu tends 
to form platelets perpendicular 
to the $\langle 100\rangle$ direction~\cite{Marceau06,Ringer00,Hono99}
and also because
according to our calculated total energies these configurations are
favored in contrast to more dense Cu clusters. However, as listed in
Table~\ref{structuretable}, in addition to these planar
configurations we have considered 3 Cu atoms as nearest neighbors to each 
other and 4 Cu atoms in a tetragonal configuration.

According to
Table~\ref{structuretable} the larger the number of Cu atoms around the
vacancy, the stronger the outward relaxation of the Cu atoms and the
inward relaxation of the Al atoms. For less than 4 Cu atoms also the Al
atoms tend to relax outward, but much less than the Cu atoms. The
relatively long positron lifetime in the case of 12 
Cu atoms is related to the strong outward relaxation, which
increases the vacancy size compared with the other defects.
The strong tendency of the Cu atoms to relax outward from the
ideal Al lattice sites is in accordance with the 
Effective Medium Theory by Jacobsen {\em et al.}\cite{Jacobsen87}
which indicates that in metallic environments Cu atoms seek to a much  
higher electron density than Al atoms. The theory explains also
why, according to our calculations, the change of the Cu atom 
configuration for a fixed number of Cu atoms does not alter
significantly the open volume of the defect and the positron lifetime.

%
\begin{table}
\caption{\label{structuretable}Relaxations and positron lifetimes for 
  vacancy-Cu complexes in the Al-Cu alloy. In the case of 4 Cu atoms,
  the planar and tetragonal configurations are considered. The ranges of 
  relaxations are given
  for both Al and Cu nearest neighbors to the vacancy. The relaxations are 
  relative to the ideal lattice positions and given in \% of the
  bond length in the strained Al. A positive (negative) value
  denotes outward (inward) relaxation.}
\begin{ruledtabular}
  \begin{tabular}{crrrr}
No.\ of Cu & Relaxation, & Relaxation, & Lifetime\\
atoms & Al (\%) & Cu (\%) & (ps)\\
\hline
0 & +1.6 & -- & 226\\
1 & +0.5 \dots +1.5 & +7.3 & 223\\
2 & +0.3 \dots +0.8 & +6.9 & 221\\
3\footnote{On the [100] plane} & --0.6 \dots +1.1 & +5.6 \dots +6.2 & 218\\
3\footnote{Nearest neighbors} & +1.1 & +5.6 \dots +6.2 & 218\\
4\footnotemark[1] & --0.9 & +5.9 & 216\\
4\footnote{Tetragonal} & --2.7 \dots +0.4 & +5.4 \dots +7.8 & 217\\
6 & --3.3 \dots --2.1 & +5.9 \dots +6.8 & 213\\
8 & --5.7 & +5.5 & 210\\
10 & --9.9 & +3.0 \dots +11.0 & 223\\
12 & -- & +7.4 & 235\\
\end{tabular}
\end{ruledtabular}
\end{table}

\subsection{\label{expfitting}Comparing the experimental Al-Cu spectra
  with calculations for model systems}

\begin{figure}
\includegraphics[width=0.9\columnwidth]{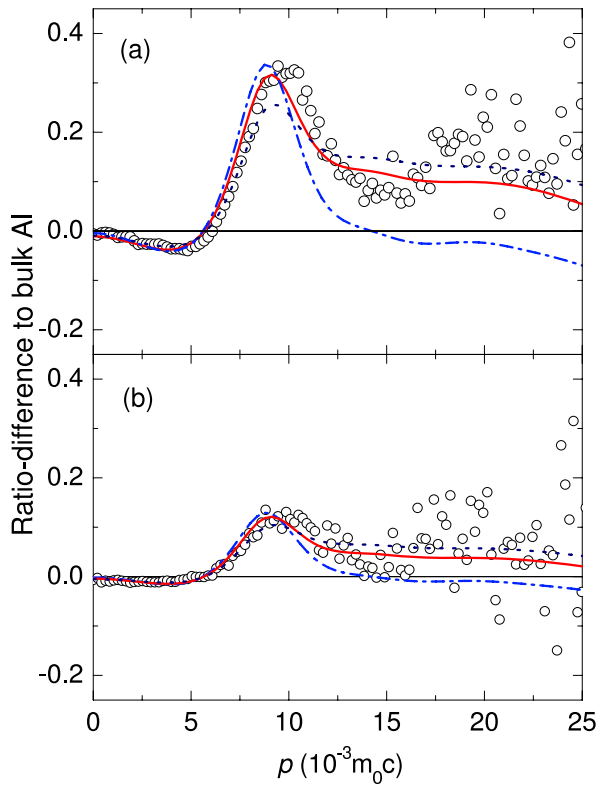}%
\caption{\label{expfit}(Color online) Experimental CDB spectrum ($\circ$)
  (RD to bulk Al) of (a) an as-quenched Al--1.1 at.~\% Cu
  sample, and (b) the same sample aged 1~min at 150~$^{\circ}$C. The 
  dash-dotted (blue), solid (red), and dotted (black)
  lines show the best one-component fits obtained for vacancy-Cu defects 
  with 2, 3, and 4
  Cu atoms, respectively.}
\end{figure}

\begin{figure*}
\includegraphics[height=0.9\textwidth,angle=-90]{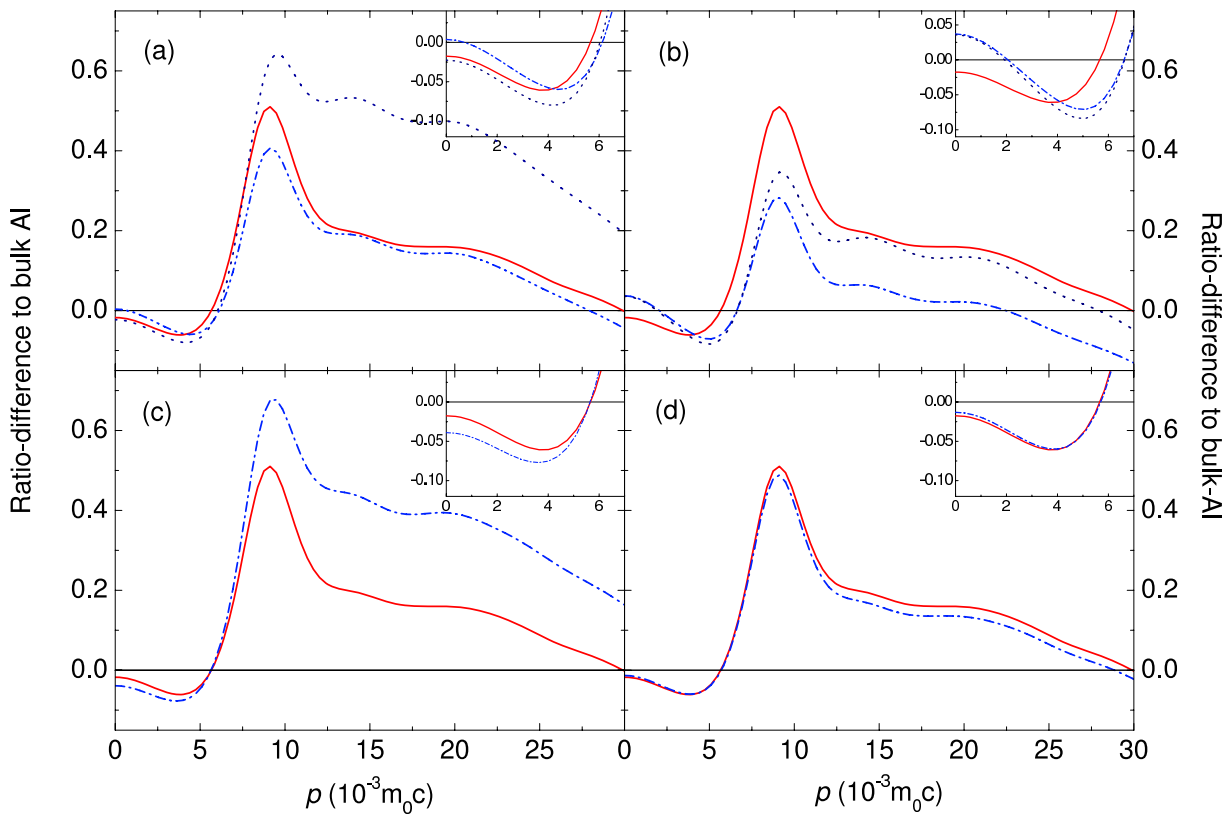}%
\caption{\label{sensitivity}(Color online) Influence of different 
  approximations on the first-principles results for Al vacancies 
  decorated with Cu atoms. In all panels the solid (red) line shows 
  the RD spectrum obtained for the Al vacancy surrounded 
  by 3 Cu atoms on the [100] plane in the strained Al lattice including ionic relaxation 
  due to the electronic structure and the localized positron state. 
  (a) The (dark blue) dash-dotted and the (blue) dotted lines show 
  the results corresponding to ideal (no relaxation) ion positions 
  and for 2 and 3 Cu atoms, respectively. 
  (b) The (dark blue) dash-dotted and the (blue) dotted lines correspond 
  to the Al lattice without strain and to 3 and 4 Cu atoms, 
  respectively. 
  (c) The (dark blue) dash-dotted curve shows the result for  
  3 Cu atoms nearest neighbors to each other. 
  (d) The (dark blue) dash-dotted curve  corresponds to the fit using the 
  linear combination of RD spectra for 2 and 4 Cu atoms
  with equal weigths.
}
\end{figure*}

We proceed by analyzing experimental CDB data for Al-Cu alloys using the 
previously described calculations for realistic model defects.
Different simulated RD spectra are scaled separately so 
that the least squares method gives the best fit with those measured  
for the as-quenched and 1-min aged Al--1.1 at.~\% Cu samples.
The experimental procedure for obtaining these spectra is explained
in Appendix. Thus, we assume first that a single defect geometry can 
represent the whole defect distribution in the sample in an average 
manner. As explained above, the scaling means the fitting of the
fraction $F$ of positrons trapped at defects. Figure~\ref{expfit}(a) shows
the results for the as-quenched
sample and simulated spectra with 2, 3,  and 4 Cu atoms. Data for
momenta up to $15\times 10^{-3}\ m_{0}c$ are
included in the fitting. The best fit is obtained with 3 Cu atoms
with the scaling corresponding to the trapping fraction of 62\%.
The fits with 2 and 4 Cu atoms are clearly worse. Figure \ref{sensitivity}(d)
below shows
that taking the linear combination of the 2 and 4 Cu atom spectra with
equal weights we obtain a fit which practically coincides with
the 3 Cu atom spectrum. Even the scaling needed remains the
same. We can therefore conclude that the sample contains a distribution
of vacancies with 3 Cu nearest neighbors on the average. 

The above analysis using the first-principles-simulated RD curves
agrees well with the analysis based on the linear combination of
measured spectra for bulk Al and for deformed Al and Cu
(see Appendix).
The fitting gives the Cu contribution of 23$\pm$1.5\%, i.e., there are
roughly 3 Cu atoms among the 12 nearest neighbors of the vacancy on the
average. The trapping fraction is found to be 73$\pm$1\% which is slightly
larger than the estimate based on the theoretical spectrum.

%

In the case of the 1-min aged sample
the experimental data points [Fig.~\ref{expfit}(b)] are quite scattered
because of the lower vacancy concentration and smaller trapping fraction.
This complicates the fitting. However, as Fig.~\ref{expfit}(b) shows we get 
a reasonable fit also in this
case when having 3 or 4 Cu atoms next to the vacancy. These fits
result in trapping fractions of 24\% or 17\%, respectively. These
numbers should be contrasted
with the experimental analysis of Appendix indicating the Cu
concentration of 33$\pm$5\%, i.e. about 4 Cu atoms around the vacancy  
and the trapping fraction of 18$\pm$1\%. However, because of
the very scattered experimental data and low trapping fraction, it is
difficult to draw accurate conclusions about the number of Cu atoms
around the vacancy. 





Now we want to get an idea about the importance of different ingredients
of our first-principles calculations on the simulated CDB RD  
spectra.
Our reference system is the Al vacancy in the strained Al lattice decorated
with 3 Cu atoms. Due to the good agreement with the experimental 
result shown in Fig.~\ref{expfit}(a) this suits well as a representative
test case.
Figure~\ref{sensitivity}(a) shows that ignoring the lattice relaxation 
relative to the ideal lattice positions increases the intensity
at high momenta and decreases the relative intensity of the peak
at around $p=9\times 10^{-3}\ m_{0}c$. This change reflects the
tendency of the Cu atoms to relax strongly outward 
(see Table~\ref{structuretable}). The reduction of the number of Cu atoms
to 2 would improve the agreement with experiment but still the shape
is worse than that of the fully relaxed 3-Cu-atom case. The importance
of using the strained Al lattice is emphasized in Figure~\ref{sensitivity}(b).
Ignoring the strain increases the intensity at low momenta indicating
a too large open volume and a too long positron lifetime. The
negative values at high momenta can be healed by increasing the
number of Cu atoms to 4 but this affects by worsening the relative
intensity of the peak. Figure~\ref{sensitivity}(c) shows that 3 Cu
atoms nearest neighbors to each other result in a large intensity increase
at high momenta with respect to the 3 Cu atoms on the same [100] plane.
This is because the dense Cu cluster creates open volume attracting
the positron density and resulting in an increased annihilation
with Cu 3$d$ electrons. However, as discussed above, the planar 
configuration is favored against the dense one due to
its lower calculated total energy and also due to the experimental
findings.

\subsection{\label{lincombination}Linear combinations of elemental CDB
  spectra: theory vs theory}

The most important goal of the present work is to test the reliability of 
the linear combination method adopted in the experimental data
analysis (see Appendix). For example, the RD spectrum $\Delta(p)$ 
for a vacancy-Cu complex in the Al-Cu alloy is fitted with the function 
\begin{eqnarray}
\label{fitfun}
\Delta^{\text{Fit}}(p) & = & F[
  C_{\text{Al}}\Delta^{\text{Vac}}_{\text{Al}}(p)+C_{\text{Cu}}\Delta^{\text{Vac}}_{\text{Cu}}(p)]\nonumber\\
& = & F[ {N_{\text{Al}} \over 12} \Delta^{\text{Vac}}_{\text{Al}}(p)+{N_{\text{Cu}} \over 12}  \Delta^{\text{Vac}}_{\text{Cu}}(p)],
\end{eqnarray}
where $\Delta^{\text{Vac}}_{\text{Al}}(p)$ and $\Delta^{\text{Vac}}_{\text{Cu}}(p)$ are
the RD spectra for vacancies in strained (i.e., deformed) Al and Cu, 
respectively.
$F$ is the fraction of trapped positrons and $N_{\text{Al}}$ and $N_{\text{Cu}}$
are the numbers of Al and Cu nearest neighbors to the vacancy, 
respectively. The Al and Cu concentrations around the vacancy
fulfill the constraint
$C_{\text{Al}} + C_{\text{Cu}} =1$,
so that there are only two free fitting parameters, e.g., $F$ and
$C_{\text{Cu}}$. Actually, $F$ should be ideally equal to unity,
but we will keep it as a fitting parameter in order to test 
the idea of the general fitting procedure.
Figure~\ref{RDelemental} shows the theoretical
elemental RD spectra we use in our fitting.

\begin{figure*}
\includegraphics[height=0.9\textwidth,angle=-90]{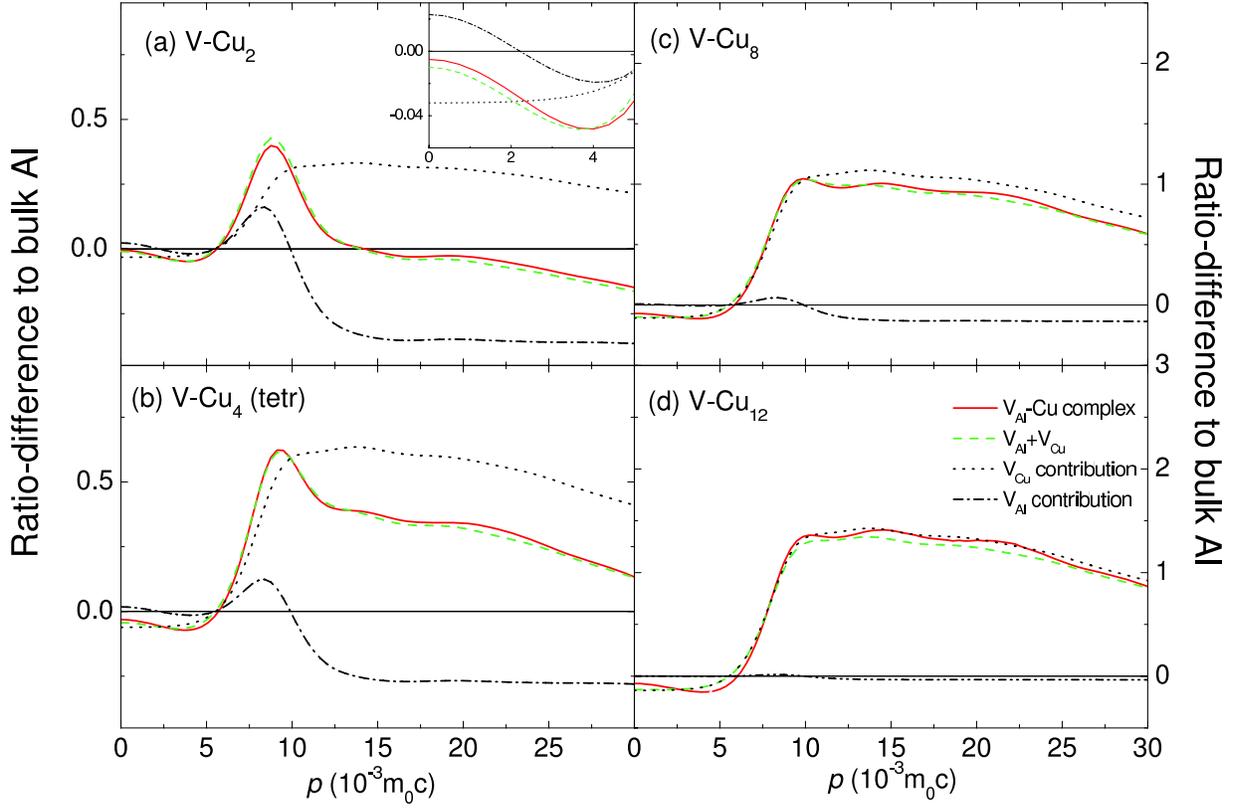}%
\caption{\label{lincombfig}(Color online) CDB RD spectra for 
  vacancy-Cu complexes in Al-Cu alloys with (a) 2, (b) 4, (c) 8, and
  (d) 12 nearest neighbor Cu atoms. The tetragonal configuration is
  used in the case of 4 Cu atoms. The solid (red) curves
  give the results of the first-principles calculations. The
  dashed (green) curves are the linear combinations [Eq.~(\ref{fitfun})]
  of the RD spectra calculated for vacancies in strained Al and Cu.
  The dash-dotted and dotted curves show the Al and Cu contributions,
  respectively. The inset in (a) shows the low-momentum region magnified.
  }
\end{figure*}

Instead of the experimental Al-Cu alloy data we use here the RD spectra 
calculated from our first-principles results for Al vacancies decorated 
with Cu atoms and fit them using the similarly calculated RD spectra 
for vacancies in strained Al and Cu. 
Figure~\ref{lincombfig} shows the simulated RD 
curves for 2, 4, 8 and 12 Cu atoms as the nearest neighbors 
to the vacancy. The linear combinations obtained with the
least squares method and their Al and Cu vacancy components are
also shown. The momentum range of 0\dots$15\times 10^{-3}\ m_{0}c$ is
included in the fitting. As in the purely experimental analysis the
fits reproduce successfully the original curves.
The decompositions show that the
peak around $8\times 10^{-3}\ m_{0}c$ in the RD spectrum
is due to the Al vacancy component and
Cu 3$d$ derived electronic states are responsible for the raise of the
intensity at high momenta.
At the low-momentum region of $0 \dots \sim 7 \times 10^{-3}\ m_{0}c$ the 
agreement between the simulated RD curve and the linear combination
worsens with increasing number of Cu atoms surrounding the
Al vacancy. This is due to the fact that the electron density at
the vacancy in Cu is larger than that at the Al vacancy surrounded
by Cu atoms.

More quantitatively, the fitting parameters $F$ and $C_{\text{Cu}}$ are
given in Table~\ref{theofit}. The fitting seems to underestimate
the Cu concentration consistently by around 10\%. 
This rather good accuracy is obtained by the help of the reduction 
of the trapping fraction from its nominal value of unity. When the 
number of Cu atoms increases from 1 to 12, $F$ decreases rather 
linearly from a value slightly less than unity to around 0.7.
The reduced trapping fraction substitutes the annihilation with
Cu by annihilation with bulk Al. This compensation is needed because 
the use of the Cu vacancy data would otherwise lead to too
strong annihilation with Cu atoms due to the fact that the Cu lattice 
constant is smaller than that for Al. The reduced trapping fraction 
decreases also the contribution $F C_{\text{Al}}$ of the Al-vacancy RD
spectrum.
This decrease is compensated by the increase of the 
ratio $C_{\text{Al}}/C_{\text{Cu}}$. But the effect of this
``secondary'' compensation is not strong
and it decreases with the decreasing number of Cu atoms so that the 
ratio $C_{\text{Al}}/C_{\text{Cu}}$ remains close to the correct
one. Thus, the Cu concentration is estimated to be only 10\% too small.

\begin{table}
\caption{\label{theofit} Fitting the RD spectra
calculated from the first-principles results for vacancy-Cu complexes 
in the Al-Cu alloy with the linear combination of calculated
RD spectra for vacancies in strained Al and Cu. 
The fitting parameters, i.e., the number $N_{\text{Cu}}$ of Cu atoms around
the Al-vacancy and the fraction of trapped positrons $F$ are defined 
in Eq.~(\ref{fitfun}).
}
\begin{ruledtabular}
  \begin{tabular}{crrrrr}
No.\ of Cu atoms & $N_{\text{Cu}}$ & $F$\\
in complex\\
\hline
1 & 1.00  & 0.967 \\
2 & 1.80 & 0.924 \\
3\footnote{On [100] plane} & 2.71 & 0.898 \\
3\footnote{Nearest neighbors} & 3.84 & 0.912 \\
4\footnotemark[1] & 3.49 & 0.842 \\
4\footnote{Tetragonal} & 3.67 & 0.868 \\
6 & 5.54 & 0.811 \\
8 & 7.38 & 0.758 \\
10 & 9.05 & 0.762 \\
12 & 10.66 & 0.671\\
\end{tabular}
\end{ruledtabular}
\end{table}

\subsection{\label{MgAg} Vacancy-solute complexes in Al-Cu-Mg and   
Al-Cu-Mg-Ag alloys}

Next we want to investigate possibilities for quantitative analysis
of vacancy-solute complexes when there are more than one type
of solute atoms. As examples we choose complexes in the technologically
important ternary Al-Cu-Mg and quarternary Al-Cu-Mg-Ag alloys.
The Al matrix around the defect complex is described by the strained
Al lattice as above for the Al-Cu alloy. The ionic and electronic
structures and the trapped positron density are calculated
with the first-principles scheme.

\begin{figure}
\includegraphics[width=0.85\columnwidth]{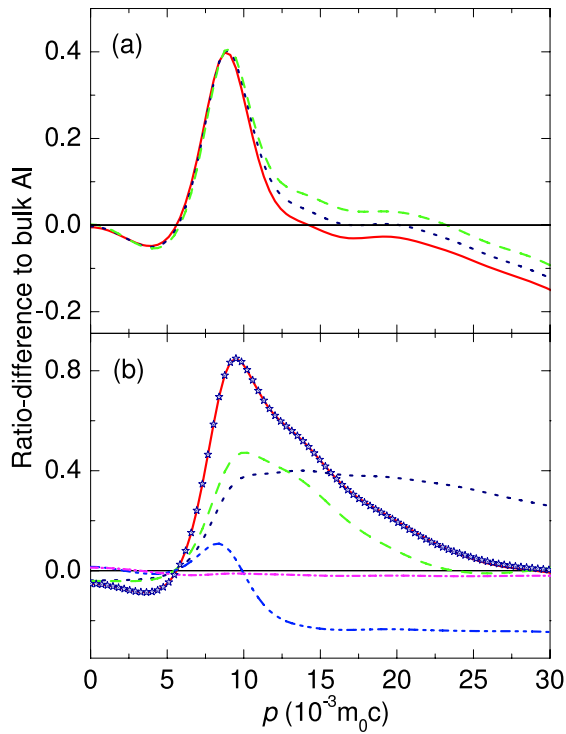}%
\caption{\label{RDternquatern}(Color online) (a) CDB RD spectra for 
  vacancy-Cu-Mg complexes in Al-Cu-Mg alloys with 2 Cu and 0, 1, and 2 Mg
  atoms denoted by solid (red), dotted (dark blue) and dashed (green) lines,
  respectively. (b) CDB RD spectrum for a vacancy surrounded by 2 Cu,
  2 Mg, and 2 Ag atoms and the best linear-combination fit denoted by
  a solid (red) line and (blue) stars, respectively. The
  components of the fit, corresponding to vacancies in strained Al,
  Cu, Mg, and Ag, are shown by (blue) dash-double dotted line, (dark
  blue) dashed line, (pink) dash-dotted line, and (green) long-dotted
  lines, respectively.}
\end{figure}

Figure~\ref{RDternquatern}(a) shows the RD spectra of vacancy-Cu-Mg complexes
when there are 2 Cu and 0, 1, and 2 Mg atoms as nearest neighbors to the
vacancy. All the solute atoms are on the same [100] plane with the two 
Cu atoms along a $\langle 110\rangle$ direction.
According to Fig.~\ref{RDelemental}(b) one would expect that the primary
effect of the increasing number of neighboring Mg atoms on the shape of the
defect RD spectrum would be to linearly increase its intensity at low
momenta. However, this is not the case. In Fig.~\ref{RDternquatern}(a) the
effect of Mg at low momenta is negligible. At high momenta where the
annihilation with Cu $3d$ derived states and Al and Mg core states dominates
in the defect spectrum the effect of Mg is to increase positron
annihilation with these states. 
First of all, Mg has a larger Wigner--Seitz radius than Al. This
means that the open volume at the vacancy decreases with the increasing
number of Mg atoms. The relaxation of Mg atoms in these systems is about
--1.3\ldots--1.2\% (minus sign denotes inward relaxation). The
positron lifetime increase per Mg atom added is 2~ps. The effect of the
decreased vacancy volume on the CDB RD spectrum is to increase
core annihilation and decrease annihilation with low-momentum valence
electrons at the vacancy. On the other hand, substituting Al by Mg
decreases the electron density at the vacancy which in turn increases
the value of the RD spectrum at low momenta. This effect, however, is
not strong enough to raise the peak at low momenta. This fact
is also seen in results of simulations in which there is only Al and
Mg around the vacancy.

The possibility of quantitative identification of Mg at the
vacancy-solute complexes is studied in Table~\ref{AnalysisAlCuMg} 
which shows the analysis results of the RD spectra calculated for
several vacancy-Cu-Mg complexes. Up to 4 solute atoms they are on the
same [100] plane. A general trend is that Cu atoms tend to relax strongly
outward from the center of the vacancy, up to $\sim 10$\% of the Al bond
length, whereas Mg atoms tend to relax slightly smaller amounts
inward. The different behaviors reflect the trends in the Wigner--Seitz  
radii. The fitting with the spectra for vacancies in elemental
strained metals is performed in the momentum range of 0\dots$15\times
10^{-3}\ m_{0}c$. According to Table~\ref{AnalysisAlCuMg} the trapping
fractions are closer to unity than
in the case of vacancy-Cu complexes in Table~\ref{theofit}. This is
because Cu has a smaller and Mg a larger Wigner--Seitz radius than
Al so that their effects tend to compensate each other. The analysis
gives rather faithfully the correct number of Cu atoms. But the number
of Mg atoms is clearly underestimated. This is because the Mg 
component in the fit, i.e., the RD spectrum for a vacancy in strained
Mg has a rather strong peak at low momenta due to the large
Wigner--Seitz radius and the low electron density. In the denser Al lattice
Mg does not decrease the electron density to the portion deduced on the
basis of the effect of the vacancy in Mg. Moreover, the effect on the
RD spectrum at low momenta is not necessary a linear function of 
the number of Mg atoms as discussed above.

Table~\ref{AnalysisAlCuMg} shows also results for
vacancy-Cu-Mg-Ag complexes. In these systems the fitting procedure
which is now extended up to $17\times 10^{-3}\ m_{0}c$
reproduces the correct trapping fraction extremely well. Also
$N_{\text{Cu}}$ and $N_{\text{Ag}}$ are described with reasonable
accuracy. Ag can be separated because due to the 4$d$ electrons its RD
spectrum has in contrast to those of Mg and Al a strong intensity at
high momenta but on the
other hand the Ag 4$d$ contribution decays much faster than the Cu
3$d$ contribution towards high momenta
[see Fig.~\ref{RDelemental}(b)]. For this reason, when analyzing
systems with Ag it is crucial to extend the momentum fitting region at least
up to $17\times 10^{-3}\ m_{0}c$ in order to be able to differentiate 
between Ag and Cu. Again, the Mg content is underestimated but the sum
$N_{\text{Al}}+N_{\text{Mg}}$ is described
well. Figure~\ref{RDternquatern}(b) shows as a typical example the
spectrum and its decomposition for the vacancy complex with 2 Cu, 2
Mg, and 2 Ag atoms.

\begin{table}
\caption{\label{AnalysisAlCuMg} Fitting the RD spectra
calculated from the first-principles results for vacancy-Cu-Mg
and vacancy-Cu-Mg-Ag complexes in the Al-Cu-Mg and Al-Cu-Mg-Ag alloys
with the linear combination of calculated RD spectra for vacancies in
strained Al, Cu, Mg, and Ag.
The fitting is done by generalizing Eq.~(\ref{fitfun}) to four
components $N_{\text{Al}}$, $N_{\text{Cu}}$, $N_{\text{Mg}}$, and
$N_{\text{Ag}}$ giving the 
number of Al, Cu, Mg and Ag atoms around the vacancy, respectively.
The fitting parameter $F$ is related to the  fraction of trapped 
positrons.
}
\begin{ruledtabular}
  \begin{tabular}{cccrrrrr}
No.\ of Cu & No.\ of Mg & No.\ of Ag &$N_{\text{Cu}}$ &
$N_{\text{Mg}}$ & $N_{\text{Ag}}$ & $F$\\
atoms & atoms & atoms &         &          &    & \\  
\hline
1 & 1 & & 1.07 & 0.32 & & 0.954\\
2 & 1 & & 1.94 & 0.58 & & 0.932\\
2 & 2 & & 2.09 & 0.99 & & 0.919\\
2 & 4 & & 2.37 & 1.69 & & 0.895\\
2 & 6 & & 2.60 & 2.51 & & 0.900\\
3 & 1 & & 2.78 & 0.89 & & 0.912\\
3\footnote{Different configuration of the nearest-neighbor solute atoms} & 1 & & 2.87 & 0.66 & & 0.906\\
3 & 3 & & 3.16 & 1.40 & & 0.894\\
4 & 4 & & 4.10 & 2.19 & & 0.868\\
5 & 2 & & 4.79 & 1.52 & & 0.882\\
2 & 2 & 2 & 2.08 & 0.71 & 2.63 & 0.963\\
2 & 4 & 2 & 2.60 & 1.40 & 1.77 & 0.980\\
4 & 2 & 2 & 3.66 & 1.12 & 2.35 & 0.975\\
3 & 1 & 1 & 3.10 & 0.53 & 0.81 & 0.994\\
3\footnotemark[1] & 1 & 1 & 3.10 & 0.74 & 0.58 & 1.01\\
3 & 1 & 2 & 2.97 & 0.51 & 1.96 & 1.00\\
3\footnotemark[1] & 1 & 2 & 2.93 & 0.55 & 2.05 & 1.01\\
\end{tabular}
\end{ruledtabular}
\end{table}

\section{\label{conclusions}Summary and conclusions}

We have performed first-principles calculations for electronic structures,
positron states, and ion positions at solute-vacancy complexes in metallic
alloys Al-Cu, Al-Mg-Cu, and Al-Mg-Cu-Ag. The ensuing positron annihilation 
characteristics, positron lifetimes and momentum distributions of annihilating
electron-positron pairs have been calculated. Since the electron and
positron densities as well as the ion positions depend on each others
they have to be determined in a self-consistent fashion. We have shown
that this step is required in order to obtain annihilation
characteristics consistent with the experimental ones.

Calculated ratio-difference curves of the Doppler broadening spectra
are compared with measured ones in order to identify quantitatively
the vacancy-solute defects, i.e., to determine the average numbers of
different solute atoms around the vacancies. An experimental spectrum
can be interpreted in terms of a spectrum calculated for a certain
defect configuration or by a linear combination of spectra for two or
more defect configurations.

By the help of the first-principles results for vacancy-solute complexes 
and vacancies in constituent metals we have tested the linear fitting 
procedure suggested by Somoza \textit{et al}.\ to find the percentages
of different
atomic species around the vacancies. With the Al-Cu, Al-Mg-Cu, and 
Al-Mg-Cu-Ag alloys as test cases we have shown that the linear combination
can resolve the average number of solute atoms whose electronic structures 
differ clearly from those of the matrix and other solute atoms (such
as Cu and Ag in our tests). Typically, an accuracy better than 0.5
\dots 1.0 atoms from the 12 nearest neighbors can be
achieved. However, adjacent atoms of the Periodic Table having similar
core electron  structures (such as Mg and Al) cannot be
distinguished.

In our test cases the most important momentum region for the linear
combination fits is above $\sim 7 \times 10^{-3}\ m_{0}c$. In this
region the behavior of the electron wave functions near the ion
cores is important leading to distinction of the different atomic 
species. The lower momenta are dominated by free-electron type behavior 
which depends mainly on the average valence electron density seen
by the positron. The introduction of solute atoms does not introduce
changes like new covalent bonds which would affect the above division
to free-electron and atomic-specific regions.
However, in order to mimic this division in linear combination fits
in a reasonable manner it is important to use the vacancy spectra,
not the bulk spectra of solute metals as components.

The linear fitting procedure works well also due to compensating
effects affecting the positron annihilation rates with different
ion cores.
The rates depend on open space seen by the
positron and this may differ between the matrix and solute metal
lattices. Compensation may occur when there are two or more different
types of solute atoms around the vacancy. Compensation may also occur
by underestimating or overestimating the fraction of trapped positrons
if the fraction is kept as a free parameter in the fit. Therefore
comparing the fitted trapping fraction
with that obtained by using the positron lifetime measurements is a valuable 
tool to check the consistency of the linear fitting procedure. 

\appendix*
\section{Details of the experiments and the experimental fitting procedure}

The experimental data presented below refer to two pairs of samples of
a well-homogenized laboratory alloy, obtained by melting commercially
pure elements. The nominal composition of the alloy was Al--1.1 at.~\%
Cu. All samples were solution treated by holding at 525$^{\circ}$C for 1~h,
then water quenched at room temperature. One pair of samples was
immediately mounted in the standard sandwich configuration with a $^{22}$Na
positron source (about $2\times 10^{5}$~Bq) sealed between Kapton foils and
brought at liquid nitrogen temperature within less than a minute since
quenching. The second pair was aged in oil bath for 1 min at 150$^{\circ}$C,
then quenched again at room temperature, mounted with the positron
source and cooled as described for the other pair. All lifetime and
CDB measurements were taken simultaneously at liquid nitrogen
temperature; set-ups and procedures were already described in
Ref.~\onlinecite{Calloni05}. The momentum
resolution was $3.69\times 10^{-3}\ m_{0}c$ (FWHM). The time resolution of the
lifetime apparatus was 250~ps (FWHM).

The analysis of the lifetime spectrum was carried out by means of the
Positronfit program~\cite{Kirkegaard89} in a single component
after subtraction of 12.5\% source component. The CDB empirical
analysis was performed by applying a modified version of the linear
combination procedure suggested by Somoza \textit{et al}.~\cite{Somoza02}

In the present version, the CDB spectrum $\rho$ was fitted with a
revised version of a linear combination (see Ref.~\onlinecite{Ferragut06})
of three reference spectra,
$\rho^{\text{bulk}}_{\text{Al}}$, $\rho^{v}_{\text{Al}}$,
$\rho^{v}_{\text{Cu}}$, measured, respectively
for: (a) bulk annealed Al (no positron trapping); (b) deformed Al
(100\% positron trapping, after correction for the contribution of
free positrons); (c) deformed Cu (100\% positron trapping, after
correction for the contribution of free positrons). The linear
combination can be written in the form
\begin{equation}\label{fittincurve}
\rho=(1-F)\rho_{\text{Al}}^{\text{bulk}}+F(C_{\text{Al}}\rho^{v}_{\text{Al}}+C_{\text{Cu}}\rho^{v}_{\text{Cu}}),
\end{equation}
where $F$ is the proposed estimate of the positron fraction trapped at
vacancy-solute clusters and $C_{\text{Al}}$, $C_{\text{Cu}}$ are the
  proposed estimates of the fractional atomic concentrations in
  contact with the vacancy.

\begin{figure}
\includegraphics[width=0.9\columnwidth]{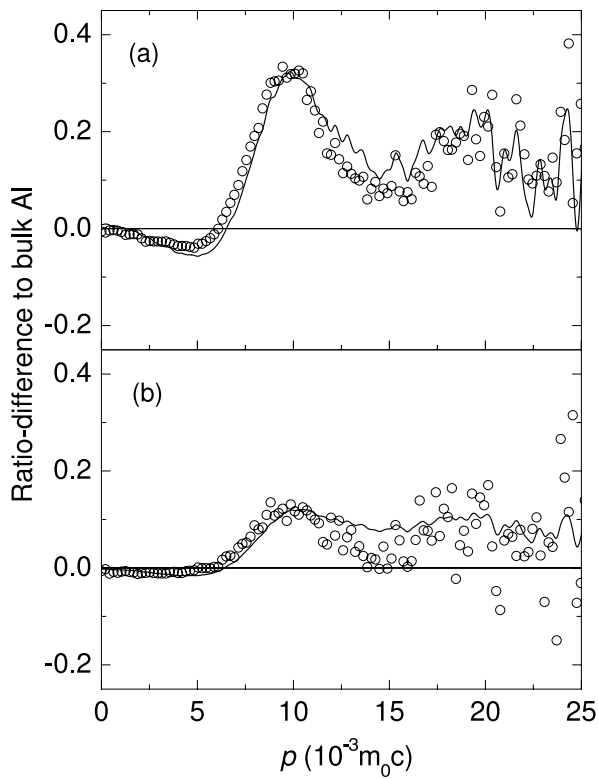}%
\caption{\label{expfig}Results of CDB measurements 
(markers) for (a)
  Al--1.1 at.~\% Cu alloy measured after quenching and (b)
  after 1~min aging at 150$^{\circ}$C. The solid lines
  show the empirical fitting curves.}
\end{figure}

The same experimental data as in Fig.~\ref{expfit} and the empirical
fitting curve [Eq.~(\ref{fittincurve})]
are shown in Fig.~\ref{expfig} versus the modulus of the momentum
component
The numerical results of the CDB analysis and
the mean positron lifetimes are reported in Table~\ref{expfittable}.

\begin{table}
\caption{\label{expfittable}Empirical fitting coefficients $F$,
  $C_{\text{Al}}$ and $C_{\text{Cu}}$ and the mean positron lifetimes $\tau$
  for the Al--1.1 at.~\% Cu alloy after quenching and
  after 1~min aging at 150$^{\circ}$C.}
\begin{ruledtabular}
  \begin{tabular}{lrrrr}
Sample & $F$ (\%) & $C_{\text{Al}}$ (\%) & $C_{\text{Cu}}$ (\%)&
$\tau$ (ps)\\
\hline
As quenched & 73$\pm$1 & 77$\pm$1.5 & 23$\pm$1.5 & 196$\pm$1\\
Aged 1 min at 150$^{\circ}$C & 18$\pm$1 & 67$\pm$5 & 33$\pm$5 & 169$\pm$1\\
\end{tabular}
\end{ruledtabular}
\end{table}

\begin{acknowledgments}
We acknowledge Prof.\ A.\ Dupasquier for his indispensable comments
and opinions during the whole research work presentented. We thank
Dr.\ T.\ E.\ M.\ Staab for informative discussions on positron
annihilation methods and alloy physics. This work has been
supported by the Academy of Finland through its Centers of Excellence
program, and by the Finnish Academy of Science and Letters, Vilho,
Yrj\"o and Kalle V\"ais\"al\"a Foundation (I.M.). P.F.\ and R.F.\
acknowledge the financial support by the Ministero dell'Istruzione,
Universit\`a e Ricerca (project PRIN 2004023079 ``Nanostructures and
light alloys'').
\end{acknowledgments}

\bibliography{Al-alloys.bib}

\end{document}